\documentclass[aps,prl,reprint,twocolumn,preprintnumbers,floatfix,nofootinbib]{revtex4-1}

\usepackage[T1]{fontenc} 

\usepackage[dvipdfm]{graphicx}
\usepackage{epstopdf}
\usepackage{mathrsfs}
\usepackage{amssymb}
\usepackage{verbatim}
\usepackage{color}
\usepackage[dvipsnames]{xcolor}
\usepackage{multirow}
\usepackage{amsmath}
\usepackage{subfig}
 \usepackage{slashed} 
\usepackage{float}
\usepackage[normalem]{ulem}
\usepackage{sidecap}
\usepackage{hyperref}
\usepackage{cancel}
\usepackage{enumerate}
\hypersetup{pdfstartview=FitV,colorlinks=true,linkcolor=blue,citecolor=red,filecolor=black,urlcolor=blue}

\newcommand*\diff{\mathop{}\!\mathrm{d}}

\newcommand{\beq}{\begin{equation}}
\newcommand{\eeq}{\end{equation}}
\newcommand{\bea}{\begin{eqnarray}}
\newcommand{\eea}{\end{eqnarray}}

\newcommand{\mrm}{\mathrm}

\usepackage[parfill]{parskip}
\begin{document}
\sloppy 

\preprint{KIAS-P20065}

\preprint{IFT-UAM/CSIC-20-160}

\vspace*{1mm}

\title{Disformal Dark Matter} 

\author{Philippe Brax$^{a}$}
\email{philippe.brax@ipht.fr}
\author{Kunio Kaneta$^{b}$}
\email{kkaneta@kias.re.kr}
\author{Yann Mambrini$^{c}$}
\email{yann.mambrini@ijclab.in2p3.fr}
\author{Mathias Pierre$^{d,e}$}
\email{mathias.pierre@uam.es}
\vspace{0.5cm}

\affiliation{$^a$Institut  de  Physique th\'eorique  , Universit\'e  Paris-Saclay,CEA,  CNRS,  F-91191  Gif-sur-Yvette  Cedex,  France}

\affiliation{$^b$School of Physics, Korea Institute for Advanced Study, Seoul 02455, Korea}

\affiliation{$^c$Universit\'e Paris-Saclay, CNRS/IN2P3, IJCLab, 91405 Orsay, France}

 \affiliation{$^d$Instituto de F\'{i}sica Te\'{o}rica (IFT) UAM-CSIC, Campus de Cantoblanco, 28049 Madrid, Spain} 
\affiliation{
$^e$
Departamento de F\'{i}sica Te\'{o}rica, Universidad Aut\'{o}noma de Madrid (UAM), Campus de Cantoblanco, 28049 Madrid, Spain}

\date{\today}

\begin{abstract} 
We generalize dark matter production to a two-metric framework whereby  the physical metric, which couples to the Standard Model (SM), is  conformally {\it and/or} disformally related to the  metric  governing the gravitational dynamics.  
We show that this setup is naturally  present in many Ultra Violet (UV) constructions, from K\"ahler moduli fields  to tensor-portal models, and from emergent gravity  to supergravity models. In this setting we study dark matter production in the early  Universe resulting  from both  scatterings off the thermal bath and the radiative decay of the inflaton. We also take into account
non-instantaneous reheating effects at the end of inflation. In this context, dark matter emerges from the production of the scalar field mediating the conformal/disformal interactions with the SM, i.e. realising a 
Feebly Interacting Matter Particle (FIMP) scenario where the suppression scale of the interaction between the scalar and the SM can be taken almost as high as the  Planck scale in the deep UV.

\vskip 1cm

\vskip 1cm

\end{abstract}

\maketitle

\setcounter{equation}{0}

\section{I. Introduction}

The presence of two geometries, governing the  gravitational dynamics and the behaviour of matter fields respectively,  is  frequent within the landscape of high-energy physics models.
 For instance, orbifolds in string theories~\cite{Callan:1988hs} or K\"ahler metrics in supergravity models~\cite{gravitino} are two popular cases where
the geometry governing
the dynamics of matter are not the same that the one governing the
gravitational structure of space-time. This is in fact quite an old idea and was already proposed
 in Nordstrom gravitational theories~\cite{Nordstrom}, Brans-Dicke's~\cite{Brans}
or Dirac's~\cite{Dirac}.
More
recently models of emergent gravity~\cite{Kiritsis:2014yqa}
 modify the metric assuming that
 gravity springs  from vector interactions generated in 
massive hidden sectors. As a consequence, the dynamical metric in  Minkowski space-time 
can be reduced to~\cite{Betzios:2020sro}
\beq
g_{\mu \nu} \equiv \eta_{\mu \nu}+ \frac{T_{\mu \nu}}{\Lambda^4} \,,
\label{Eq:metric}
\eeq
\noindent
where $T_{\mu\nu}$ is a linear combination of  energy--momentum tensors of hidden sector particles and $\Lambda$ the scale beyond which the theory breaks down\footnote{Typically the mass scale of the hidden sector.} [the Beyond the Standard Model (BSM) scale]. 
Phenomenological consequences of emergent gravity in dark matter phenomenology and for the early Universe evolution have been studied in~\cite{Anastasopoulos:2020gbu}. Other possibilities
are common in the supergravity framework and involve 
the presence of moduli fields,
inducing new couplings to the Standard
Model of the type~\cite{Chowdhury:2018tzw}
\beq
{\cal L}_T^{\rm SM} \supset Z_H |D_\mu H|^2 \,
\eeq
\noindent
in the Higgs sector
with $Z_H = 1 + \frac{1}{\Lambda} ~t$, $t$ being the real part of moduli fields. This setup can also be considered as a modification of the physical geometry, dependent on the moduli fields, especially their stabilized values at the minimum of the K\"ahler potential. Finally one finds similar examples in high-scale SUSY models~\cite{Benakli:2017whb}. Indeed, the minimal coupling
of a gravitino\footnote{The spin-$\frac{3}{2}$ superpartner of the graviton.} to the 
Standard Model,  whose longitudinal mode is the Goldstino denoted by $\Psi_{3/2}$, is built by first defining a vierbein~\cite{va}
\begin{equation}
e_\mu^\alpha \ = \ \delta_\mu^\alpha -\frac{i}{2 F^2}
\left( \partial_\mu \bar \Psi_{3/2} \gamma^\alpha { \Psi_{3/2}} +  \bar \Psi_{3/2} \gamma^\alpha \partial_\mu {\Psi_{3/2}} \right)
\ , 
\label{va1}
\end{equation}

\noindent
$\sqrt{F}$ being related to the SUSY breaking scale\footnote{In this case, we can identify $\Lambda$
to $\sqrt{F}=\sqrt{m_{3/2} M_P}$, $m_{3/2}$ being the gravitino mass.}.
This enters clearly in the category of models where the Standard 
Model fields interact with the gravitino (Goldstino) through its presence in the {\it physical}\footnote{The physical metric is also called the Jordan metric and is the one coupled to the SM fields. The Einstein metric is the one governing the dynamics of space-time and enters in the normalised Einstein-Hilbert term of General Relativity. }metric $g^{\mu \nu} = e^\mu_a e^\nu_b \eta^{ab}$. 

As we have just seen, several constructions include two geometries, and the relation between the gravitational metric and the physical one implies modifications of the dynamics and the phenomenology of SM fields by the introduction of new couplings, new interactions and new fields.
Some time ago, it was proposed~\cite{Bekenstein:1992pj} to generalize this in a unique description where all types of dynamical metrics can respect some basic principles. The generalization consists in considering two metrics which are  not just conformally related. This more natural relationship involves  Finslerian geometry rather than Riemannian geometry.
Finslerian geometry is the most general geometry where the squared relativistic interval $\diff s^2$ is homogeneous of {\it second degree} in
the element $\diff x$, in other words,
\beq
\diff s^2(x,\mu \diff x) = \mu^2 \diff s^2(x,\diff x)\,.  
\label{Eq:ds2}
\eeq
\noindent
By introducing a scalar field $\phi$, one can
define a generic function $F(\phi, X, Y)$ according to 
\beq
\diff s^2 = \tilde g_{\mu \nu} \diff  x^\mu \diff x^\nu 
=g_{\mu \nu} \diff x^\mu \diff x^\nu F(\phi,X,Y)\,,
\label{Eq:ds2bis}
\eeq
\noindent
where 
\begin{equation}
    X = g^{\alpha \beta}\partial_\alpha \phi \partial_\beta \phi
\quad \text{ and } \quad
Y = \frac{\partial_\alpha \phi \diff x^\alpha \partial_\beta \phi \diff x^\beta}{g_{\alpha \beta} \diff x^\alpha \diff x^\beta}.
\end{equation}
\noindent
It can be shown that
the Finslerian condition Eq.~(\ref{Eq:ds2})
can be satisfied by
\beq
F=C(\phi, X) + D (\phi, X) Y\,,
\label{Eq:F}
\eeq
\noindent
with $C > 0$ and $D>0$ to {preserve  the signature 
$(+,-,-,-)$ 
and respect causality~\cite{Bek2}.
Combining Eq.~(\ref{Eq:ds2bis}) with (\ref{Eq:F})
we obtain for the {\it physical} metric $\tilde g_{\mu \nu}$
\beq
\tilde g_{\mu \nu} = 
C(\phi,X) g_{\mu \nu} 
+ D(\phi,X) \partial_\mu \phi \partial_\nu \phi\,.
\label{Eq:gtilde}
\eeq
The expression (\ref{Eq:gtilde}) contains not only the classical 
{\it conformal} transformation induced by $C$ between the two metrics, but also the possibility for a {\it disformal}
transformation through the coefficient $D$, disformal in the sense that the space-time structure is stretched differently in each direction proportionally
to $\partial_i \phi$ in the $i$th-direction.
As expected, if $\phi$ is a constant, i.e a  homogeneous and isotropic field, 
both metrics are related by a simple conformal transformation.
If not ($D \neq 0$), $\phi$ is interacting with the matter fields through their kinetic terms. Notice that the metric $\tilde g_{\mu\nu}$ can also be inferred by requiring general covariance and the absence of derivatives of order larger than two. The latter requirement follows from the generic appearance of ghosts in theories with higher order derivatives.
Disformally related metrics have been widely used in the cosmological, gravitational and recently particle physics contexts \cite{Zumalacarregui:2010wj,Karwan:2016dfm,Brax:2014vva,Brax:2015hma,Brax:2016did,Brax:2019tcy}. For instance, the authors of~\cite{Dusoye:2020wom} have recently given to $\phi$ the role of the quintessence field of dark energy and 
analysed the parameter space defined by ($C$, $D$) which is cosmologically allowed. Similarly in the gravitational context, the authors of~\cite{Anson:2020trg} have constructed disformal versions of the Kerr space-time. 
 We will use disformally related metrics to induce dark matter production. In this setting,   we will  unravel how the phenomenology of the early Universe and the appearance of dark matter  
could be modified by the introduction of disformal coefficients. We will also restrict the corresponding parameter space from late- time observables.
In particular, we will show that a FIMP scenario for dark matter production can be naturally realised with a disformally coupled scalar field $\phi$ to the standard model.

Recently, ~\cite{Trojanowski:2020xza} 
 proposed that $\phi$ could play the role of a portal between a WIMP dark sector and the visible sector. 
However, the WIMP paradigm is nowadays under high scrutiny due to the lack of observed signal, especially in direct detection experiments. In~\cite{Trojanowski:2020xza}
the direct detection constraints were not taken into account, but could drastically reduce the
allowed parameter space, especially for dark matter masses below 100 GeV.
Indeed, the more recent measurements
exclude proton-WIMP cross section $\sigma \gtrsim 10^{-46} \rm{cm^2}$ for a 100 GeV 
dark matter mass~\cite{XENON,LUX,PANDAX},
which is more than six orders of magnitude
below the cross section for the vanilla
models of weakly interacting particles~\cite{Higgsportal,Zportal}. 
The simplest cosmologically viable extensions of the Standard Model reproducing the relic abundance observed by Planck experiment~\cite{planck} require to invoke a new physics scale $\simeq 5$ TeV~\cite{Arcadi:2017kky}
which will  be probed in the next
generation of experiments~\cite{Aalbers:2016jon}. If no signal is seen, this BSM scale will be pushed even further well above 50 TeV. 

However, relaxing the requirement of thermal equilibrium  between the dark sector and the primordial plasma
opens a completely new field of research~\cite{fimp,Bernal:2017kxu}. A Feebly Interacting Massive Particle (or Freeze-In Massive Particle, FIMP)
couples too weakly with the Standard Model bath to reach  thermal equilibrium in the early Universe. Such seclusion appears naturally in models where the mediator is very heavy, e.g. 
$Z'$ of unified theories ~\cite{Bhattacharyya:2018evo,SO10}, massive spin-2 particles~\cite{Bernal:2018qlk}, moduli~\cite{Chowdhury:2018tzw}, inflaton--like portals~\cite{HighlyDecoupled} or in the Kaluza-Klein theory framework~\cite{Bernal:2020fvw}. Another possibility is
to consider  theories where the couplings are
reduced by a mass parameter of the order of the  Planck mass scale $M_P$, as in supergravity\footnote{We will use throughout our work $M_P=(8\pi G_N)^{-1}\approx 2.4 \times 10^{18}$ GeV for the reduced Planck mass.}~\cite{gravitino} or a combination of the supersymmetry breaking scale and the Planck mass  in High Scale SUSY scenarios~\cite{Benakli:2017whb}. In all these cases, the temperature dependence of the production rate renders the physics in the earliest stages of the Universe more complex than the vanilla reheating scenarios described in~\cite{Giudice:2000ex}. 
Non-instantaneous thermalization~\cite{Reheating}
or non-instantaneous reheating~\cite{Garcia:2017tuj,Elahi:2014fsa} modify drastically 
the distribution function and/or the production rate of 
particles in the Standard Model plasma, making the dark
matter density calculation more complex. Considering quantum effects to the inflaton decay~\cite{Kaneta:2019zgw} or the possibility of non-standard inflaton potentials~\cite{Garcia:2020eof,Bernalbis}, show that the study
of physics at the end of the coherent oscillation 
stage at the end of inflation should be treated with care.

In, this work, we propose to consider scenarios where the dark matter is composed of the  field $\phi$ which defines
the physical metric. Indeed, a quick look at Eq.~(\ref{Eq:gtilde}) shows that the disformal term is decreasing in magnitude with the BSM scale
$\Lambda$, above which the dynamical version of the  metric breaks down anyway. This scale suppression should seclude 
$\phi$ sufficiently from the Standard Model plasma to suppress the production of $\phi$ and
making it a perfect FIMP candidate.
Moreover, the form of the metric, dictated by the consistency conditions (conservation of the signature and causality) implies a discrete $\mathbb{Z}_2$ symmetry which ensures the stability of $\phi$.

This paper is organized as follows. After a description of our models and the expression of the couplings generated by a disformal metric in section II, we compute the dark matter abundance in section III and its phenomenological consequences before concluding. 
Throughout this work, we use a natural system of units in which $k_B=\hbar=c=1$. All quantities with dimension of energy are expressed in GeV when units are not specified.

\section{II. The setup}

\subsection{The Lagrangian}

Supposing that the action is divided into a part driven
by the {\it geometrical} (gravitational) metric $g$, whereas
 matter follows the geodesics of a {\it physical} (dynamical)
metric $\tilde g$, we can write 
\beq
S = S^g_\phi + S^{\tilde g}_m = \int \diff ^4 x \sqrt{-g} {\cal L}_\phi(g, \phi) 
+ \int \diff ^4 x \sqrt{-\tilde g}{\cal L}_m(\tilde g, h ) \,,
\eeq
\noindent
with
\bea
&&
{\cal L}_\phi = \frac{1}{2}g_{\mu \nu} \partial^\mu \phi \partial^\nu \phi
- \frac{1}{2} m^2_\phi \phi^2,
\eea
\noindent
the matter Lagrangian ${\cal L}_m(\tilde g, h ) $ being the SM Lagrangian expressed in term of the metric $\tilde g$. For instance, considering one real scalar SM degree of freedom (denoted by $h$) for simplicity, gives
\begin{equation}
    {\cal L}_m = \frac{1}{2}\tilde g_{\mu \nu} \partial^\mu h \partial^\nu h
- V(h)\,.
\label{Eq:lagrangians}
\end{equation}
\noindent
\noindent
In the literature, the scalar $\phi$ is often directly or indirectly related to dark energy, 
or represents the quintessences field, and a shift symmetry
$\phi \rightarrow \phi + c$ is then imposed to avoid dangerous mass terms.
In our case, as we want to be as generic as possible, we do not impose this symmetry. Concerning the matter fields, we restrict ourselves to a singlet-like scalar $h$ field, representing the Higgs boson, to simplify the equations and explanations. Of course, the complete particle content of the Standard Model is considered for our numerical results. For the same reason, the partial derivatives in Eq.~(\ref{Eq:lagrangians}) should be understood as covariant derivatives. However, as discussed further on, the covariant part of the derivatives do not contribute significantly to the DM production and therefore are omitted for the sake of simplicity. By expanding the physical metric $\tilde g_{\mu \nu}$ in 
terms of the geometrical metric $g _{\mu \nu}$ and a small deviation $\delta \tilde g_{\mu \nu} \ll g_{\mu \nu}$, justified by the fact that we consider processes occurring at energies much below the BSM scale $\Lambda$, the matter action $S^{\tilde g}_m$
can be expressed in the Einstein frame as
\beq
S_m^{\tilde g} = S_m^g 
- \frac{1}{2}\int \diff ^4x \sqrt{-g}  ~\delta \tilde g_{\mu \nu}~T_m^{\mu \nu} 
\equiv S_m^g + \int \diff ^4x \sqrt{-g} {\cal L}_\text{int}.
\eeq
\noindent
at lowest order in $\delta \tilde g_{\mu \nu} / g_{\mu \nu}\ll1$. With $T^m_{\mu \nu}$ the energy-momentum tensor of matter fields, from Eq.~(\ref{Eq:gtilde}) we have
\beq
\delta \tilde g_{\mu \nu} = \big(C(\phi,X) -1\big) g_{\mu \nu} 
+ D(\phi,X)  \partial_\mu \phi \partial_\nu \phi\,,
\eeq
\noindent
giving
\beq
{\cal L}_\text{int}= \frac{1}{2}\big(1-C(\phi,X) \big)(T_m)^\mu_\mu 
- \frac{1}{2}D(\phi,X) \partial_\mu \phi \partial_\nu \phi T^{\mu \nu}_m\,,
\eeq
\noindent
where $T_{\mu \nu}^m$ can be expressed as
\begin{equation}
    T_{\mu \nu}^m =\sum_{i=0,1/2,1}T^i_{\mu\nu}\,,
\end{equation}
\noindent
where the sum is performed over all SM particles of spin $i$, whose corresponding energy-momentum tensors are given by
\bea
T^0_{\mu \nu} &=& \partial_\mu h~\partial_\nu h 
-g_{\mu \nu}\left[ \dfrac{1}{2} \partial^\alpha h ~\partial_\alpha h -V(h)\right]\, ,
\nonumber
\\
T^{1/2}_{\mu \nu} &=& \frac{i}{4} \left[ \bar f  \gamma_\mu \partial_\nu f - \partial_\mu \bar f \gamma_\nu f + (\mu \leftrightarrow \nu )\right]  -i g_{\mu \nu } \bar f \slashed{\partial} f \nonumber\, ,
\\
T^{1}_{\mu \nu} & = & \frac{1}{2} \left[ F_\mu^\alpha F_{\nu \alpha} + F_\nu^\alpha F_{\mu \alpha} - \frac{1}{2} g_{\mu \nu} F^{\alpha \beta} F_{\alpha \beta} \right]\, ,
\eea

\noindent 
for scalar ($h$), fermionic ($f$) and vectorial ($A_\mu$) matter fields respectively.
$F_{\mu \nu}=\partial_\mu A_\nu - \partial_\nu A_\mu$ is the field strength
of the spin-1 field whereas $V(h)$ represents the scalar potential.
Masses of various SM states are discarded as the typical temperatures 
involved in early Universe processes are much above the electroweak scale. 
Terms of the form
$g^{\mu \nu} V(h)$  can be discarded in the scalar energy-momentum tensor. Indeed, as discussed below, such terms correspond to processes involving a higher number of SM particles and/or suppressed by additional SM couplings, compared to processes relevant for the DM production. Moreover, the term $i g_{\mu \nu } \bar f \slashed{\partial} f$ in the fermionic energy-momentum tensor vanishes for on-shell states. In addition, the trace of the energy-momentum tensor $(T^i)^\mu_{\mu}$ vanishes for $i=1/2$ and $i=1$ but not for $i=0$. This is due to the fact that the energy-momentum tensor for massless states acquires a conformal symmetry in four dimensions for fermions and vectors but only in 2 dimensions for scalars. As an example, the interaction term between our DM candidate $\phi$ and one SM real scalar degree of freedom $h$ is given by %
\begin{align}
\mathcal{L}_\text{int} =  & -\dfrac{1}{2}  D(\phi,X  ) \big(   \partial_\mu \phi \partial_\nu \phi \partial^\mu h \partial^\nu h - \dfrac{1}{2} \partial_\mu \phi \partial^\mu \phi  \partial_\nu h \partial^\nu h \big) \nonumber \,, \\&+
\dfrac{1}{2} \big(C(\phi,X)-1\big) \big(  \partial_\mu h  \partial^\mu h \big) \,  .
\label{Eq:lagrangian}
\end{align}

\noindent
Notice that we have not yet made explicit the functions $C(\phi,X)$ and $D(\phi,X)$. The only assumption made was that these functions yield a small $\delta \tilde g_{\mu \nu} / g_{\mu \nu}\ll 1$. The literature is replete with clever propositions, ranging from 
invoking shift symmetries as in~\cite{Trojanowski:2020xza} with a quintessence point of view where $C$ and $D$ depend only on $X$
\cite{Brax:2016kin}, 
to supposing 
constant $X=\partial^\mu \phi \partial_\mu \phi$ in studies of 
Kerr Black Holes~\cite{Anson:2020trg}.
Other popular examples are Horndeski theories which transform 
into themselves under special disformal transformations of the metric
when $C$ and $D$ depends only on $\phi$ and not on $X$~\cite{Achour:2016rkg}. 
In this context, the resulting theories form almost the most general class of ghost-free 
scalar-tensor field theories.
$C$ and $D$ can also be considered as dependent on $\phi$ only
with expression of the type given by
\beq
C(\phi) = e^{c \frac{\phi}{M_P}}\,,~~~~~~~~
D(\phi)=\frac{d}{\Lambda^4}e^{\tilde c \frac{\phi}{M_P}}.
\label{Eq:candd}
\eeq
\noindent
In our case, we propose to expand $C$ and $D$ around $|\phi|^2$
(to ensure their positivity) which means
\bea
&&
C(\phi) \simeq 1+ c^2 \frac{|\phi|^2}{M_P^2} 
+ c_X \frac{|\partial^\mu \phi \partial_\mu \phi|}{M_P^4},
\label{Eq:conformalcoupling}
\\
&&
D(\phi) \simeq \frac{d}{\Lambda^4} + \frac{d}{\Lambda^4}
~\tilde c^2 \frac{|\phi|^2}{M_P^2}.
\eea
\noindent
Considering processes at energies much below $M_P$, it is reasonable
to stop the expansion to the first term as a first approximation.
Following (\ref{Eq:metric}), notice that such terms emerge from the coupling to a scalar of mass $m$ with
\begin{equation}
T_{\mu\nu}= \partial_\mu \phi \partial_\nu\phi 
-g_{\mu\nu} \left(\frac{1}{2} \partial^\alpha\phi\partial_\alpha \phi - \dfrac{m^2}{2} \phi^2\right)\,,
\end{equation}
where we identify $\Lambda=M_P$, $d=1$, $\tilde c=0$, $c_X=-1/2$ and $c=\frac{m}{\sqrt 2 M_p}$. In the following we will leave these parameters free in a phenomenological way.

\section{III. Dark matter phenomenology}

\subsection{Disformal production process}

Now that the Lagrangian is defined, one can investigate the DM production processes through scattering off Standard Model particles. As we commented in the previous section, we will focus our 
analysis on the case of a Standard Model bath composed of a real scalars $h$, whereas the numerical calculations will be done with the complete
set of SM particles. The corresponding Feynman diagram is shown in 
Fig.~\ref{Fig:feynman}
\begin{figure}[ht]
\centering
\includegraphics[width=2.in]{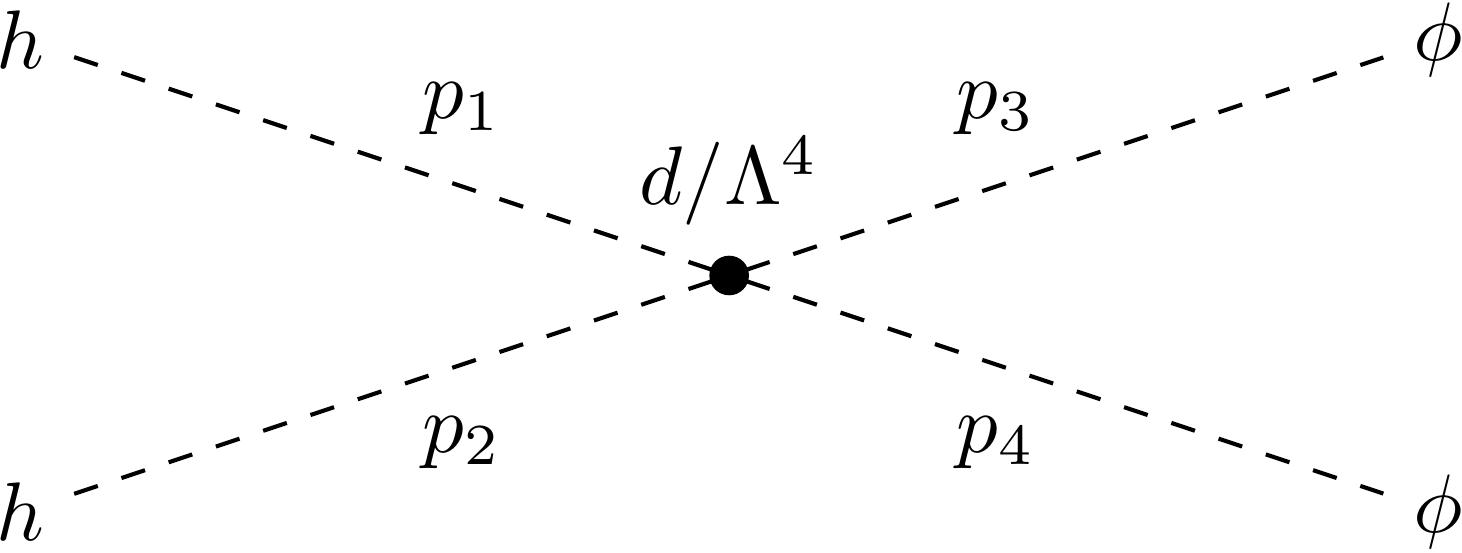}
\caption{\em \small Example of scattering processes leading to the increase of population for the dark matter candidate $\phi$.  $p_{1,2}$ and $p_{3,4}$ denote the momenta of the incoming and outgoing particles respectively.}
\label{Fig:feynman}
\end{figure}
The production rate associated to this process $1+2 \rightarrow 3+4$ where 1, 2 denote particles of the SM and 3, 4 the $\phi$
states as represented in Fig.~\ref{Fig:feynman} 
at a temperature $T$ can be written (see appendix for details and notations, as well as~\cite{Benakli:2017whb})
\beq
R(T) = \frac{1}{1024 \pi^6}\int f_1 f_2 E_1 \diff E_1 E_2 \diff E_2 \diff \cos \theta_{12}\int |{\cal M}|^2 \diff \Omega_{13}.
\nonumber
\eeq
\noindent
where $E_i$ denote the energy of particle $i=1,2,3,4$ and
\beq
f_{i}= \frac{1}{e^{E_i/T}\pm 1}\,
\eeq
represents the (thermal) distribution of the incoming particles\footnote{We consider instant thermalization in this work. For more details regarding the non-instantaneous thermalization framework, we redirect the reader to Ref.\,\cite{Reheating}.}.
Using the Lagrangian of Eq.~(\ref{Eq:lagrangian}), the scattering amplitude ${\cal M}$ can be written
\beq
{\cal M} = \frac{d}{2\Lambda^4}t(s+t)\,,
\eeq
\noindent
where $s$ and $t$ are the  Mandelstam variables.
\noindent
We then obtain for the production rate from the scalar
scatterers
\beq
R_0 = \frac{8d^2\pi^7}{297675}\frac{T^{12}}{\Lambda^8}\,,
\eeq
\noindent
Including the complete spectrum of the thermalized Standard Model species, i.e. production rates from fermions $R_{1/2}$ and vectors $R_{1}$, one obtains the total rate as
\beq
R(T) = 4R_0 +45 R_{1/2}+12R_1 \equiv \beta_d \frac{T^{12}}{\Lambda^8} \,,
\label{Eq:totalrate}
\eeq
\noindent
with $\beta_d \simeq 4\,d^2$. The exact expression for the rate is given in Eq.~(\ref{eq:betad}) and more details regarding the derivation can be found in the Appendices.

Once we know the production rate $R(T)$, the relic abundance computation is relatively straightforward. One needs to solve the integrated} Boltzmann equation
\beq
\frac{\diff n_\phi}{\diff t} + 3 H n_\phi ~= ~ R(t)\,,
\label{Eq:boltzmann0}
\eeq
\noindent
where $R(t)$ denotes the production rate of dark matter (per unit volume per unit time), or in term of temperature supposing an instantaneous thermalization,
\beq
\frac{\diff  Y_{\phi}}{\diff T} ~ = ~ - \frac{R(T)}{H (T)~T^4}\,,
\label{Eq:boltzmann}
\eeq
\noindent
with $Y_{\phi} \equiv n_{\phi}/T^3$, $H(T)= \sqrt{\frac{g_T \pi^2}{90}} \frac{T^2}{M_P}$, $g_T$ being the effective number of relativistic degrees of freedom at the temperature $T$. Solving the Boltzmann equation when 
\beq
R(T)=\beta\frac{T^{12}}{\Lambda^8},
\noindent
\eeq
\noindent
with $\beta$ a given constant gives for $T\ll T_{\rm RH}$
\begin{equation}
Y^{\rm scat }_\phi(T) \equiv Y^{\rm scat }_\phi=\sqrt{\frac{90}{g_T \pi^2}} \frac{\beta M_P}{7 \Lambda^8} T_\text{RH}^7\,,
\label{Eq:Ytrh}
\end{equation}
where $Y^{\rm scat }_\phi$ is constant for $T\ll T_{\rm RH}$. We assumed a vanishing dark matter density prior to reheating. We have {\it defined} the reheating temperature by the condition $\rho_\Phi(T_\text{RH})$ = $\rho_R(T_\text{RH})$, 
 ($\Phi$ being the inflaton field) in other words, when radiation and inflaton densities equilibrate. Notice that different definitions of the reheating temperature can lead to slightly different results, but differing never more than by factors of the order of unity as is shown for instance in~\cite{Garcia:2020eof}. 
 
 The dark matter number density reaches its maximum almost immediately after the reheating process for a temperature of $(\frac{3}{10})^{1/7} T_{\text{RH}}$ and decreases at lower temperature with a constant $n_\phi/T^3\equiv Y^{\rm scat}$.
The present relic abundance, at $T=T_0$, is given by
\beq
\Omega_\phi^{\rm scat} h^2 =\frac{n^{\rm scat}_\phi(T_0) m_\phi}{\rho_c^0/h^2}
\simeq 1.6 \times 10^{8} Y^{\rm scat }_\phi
\left(\frac{g_0}{g_\text{RH}} \right)
\left(\frac{m_\phi}{1~\mrm{GeV}} \right)\,,
\label{Eq:omega}
\eeq
\noindent 
where $\rho_c^0/h^2 = 1.05 \times 10^{-5} ~\mrm{GeV \,cm^{-3}}$ is the present critical density and $g_i$ is the effective number of degrees of freedom at temperature\footnote{With $g_0=3.91$, $g_\text{RH}=106.75$ for reheating temperatures larger than the top-quark mass $T_\text{RH}>m_t$ in the Standard Model.}  $T_i$. From Eq.~(\ref{Eq:Ytrh}) we can compute the relic abundance of the $\phi$ field produced by scattering processes
\beq
\Omega_\phi^{\rm scat}h^2 \simeq 2.7 \times 10^8 ~\beta \left(
\frac{T_{\rm RH}^7 M_P}{g_{\rm RH}^{3/2}\Lambda^8}
\right)
\left( \frac{m_\phi}{1~\rm{GeV}} \right)\,,
\eeq
\noindent
which gives, in the case of the disformal coupling, replacing the value
of $\beta$ by $\beta_d$ as computed in Eq.~(\ref{Eq:totalrate})
\beq
\frac{\Omega^{\rm scat}_{\phi,d} h^2}{0.1} =  \left(\frac{d^2}{0.4}\right)
\left(\frac{T_\text{RH}}{10^{11}}\right)^7
\left(\frac{10^{14}}{\Lambda} \right)^8 
\left( \frac{m_\phi}{10^{10}} \right) \,.
\label{Eq:omegatotscat}
\eeq
\noindent
All quantities with dimension of energy are expressed in GeV when units are not specified. We notice that, as we could have expected, the large suppression factor $\frac{d^2}{\Lambda^8}$ implies to focus on  heavy dark matter candidates due to its very feeble production in the early stage of the reheating process. We can also extract an {\it upper} bound
on $\Lambda$ from the condition $m_\phi \lesssim T_\text{RH}$ for the production
to be kinematically allowed. We then obtain
\beq
\Lambda \lesssim 10^3 ~  d^{\frac{1}{4}}~  T_{\rm RH}\,,
\eeq
\noindent
$d$ being by definition of the order of unity\footnote{Much smaller (or larger)
values of $d$ can always be absorbed in the definition of the BSM scale 
$\Lambda$.}.
This condition reflects the difficulty of producing $\phi$ in the earliest stage
of the Universe. Planck mass couplings for instance would not be sufficient
to produce dark matter with the right abundance, the majority of the reheating 
models predicting  $T_{\rm RH} \lesssim 10^{12}$ GeV~\cite{Garcia:2020eof}.

\subsection{Production from inflaton decay}
\begin{figure}[ht]
\centering
\includegraphics[width=2.5in]{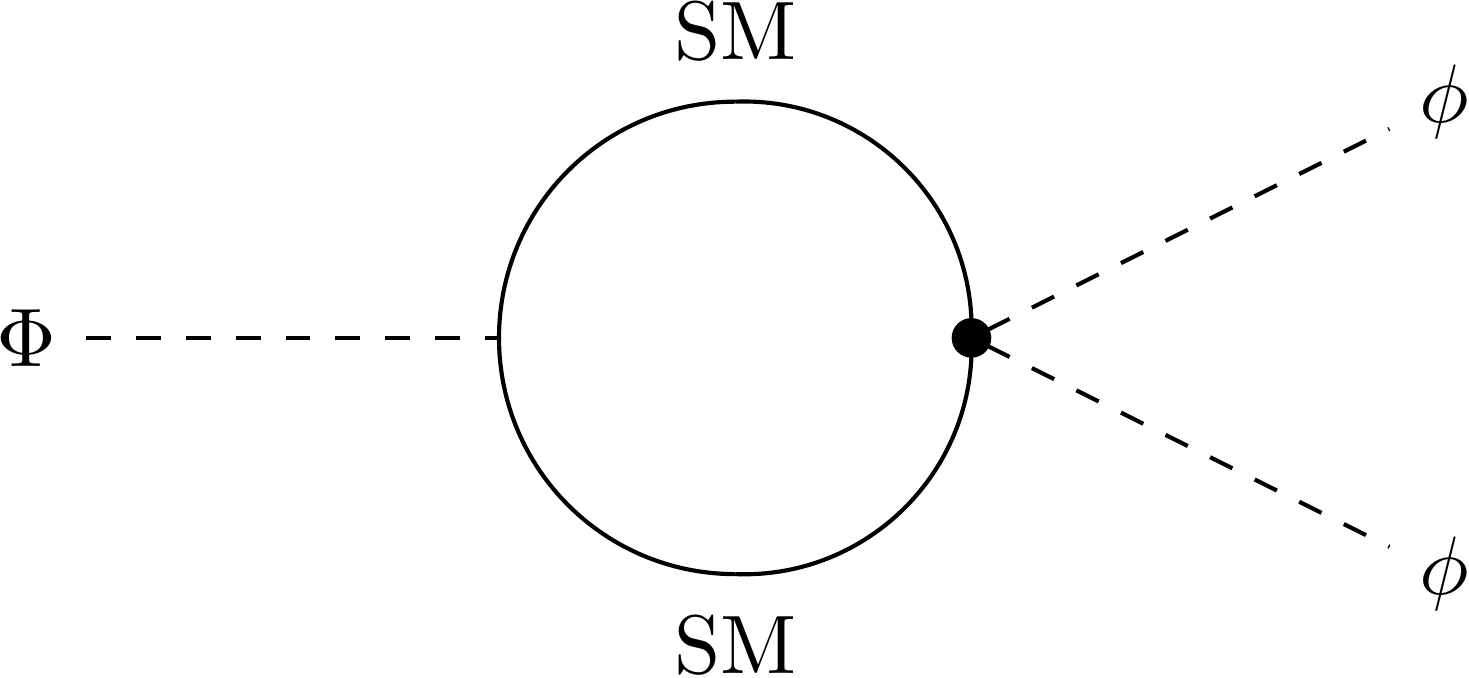}
\caption{\em \small Production of dark matter through inflaton decay, induced by a loop of SM particles.}
\label{Fig:feynmanbis}
\end{figure}
It was shown in~\cite{Kaneta:2019zgw} that if dark matter is produced by 
scattering one cannot avoid the concomitant direct production of dark matter through the loop-induced inflaton decay, as depicted in Fig.~\ref{Fig:feynmanbis}.
The minimal way to couple the Standard Model sector
to the inflaton field $\Phi$ and to realize the reheating process is via  the Higgs $SU(2)_L$ doublet $H$:
\begin{equation}
{\cal L}_\Phi \, =\, \mu_\Phi \Phi |H|^2 \, = \, \dfrac{\mu_\Phi}{2}  \Phi \sum_{i=1}^4 h_i^2\,,
\end{equation}
where $h_i$, with $i=1,2,3,4$, denote the four real scalar degrees of freedom of the Higgs doublet above the electroweak symmetry breaking scale. The decay width of the inflaton into these fundamental scalars is given by
\begin{equation}
\Gamma^\Phi_{HH}\,= \,4 \Gamma^\Phi_{h_i h_i}\,=\, \frac{\mu_\Phi^2}{8 \pi m_\Phi}\,,
\label{Eq:inflaton-higgs}
\end{equation}
\noindent
The loop-induced decay width of the inflaton to a DM pair, whose corresponding diagram is depicted in Fig.~\ref{Fig:feynmanbis}, can be expressed as
\bea
&&
\Gamma^{\Phi,d}_{\phi\phi} = 
\frac{d^2 |\frac{5}{18}-i \frac{\pi}{6}|^2}{8 \pi (16 \pi^2)^2}
\frac{\mu_\Phi^2 m_\Phi^7}{\Lambda^8}\,,
\label{Eq:widths}
\eea
\noindent
which gives for the number density of $\phi$
\bea
&&
n^\text{dec}_\phi(T_\text{RH})= B_R \frac{\rho_\Phi(T_\text{RH})}{m_\Phi}
= B_R
\left(\frac{g_\text{RH} \pi^2}{30} \right)
\frac{T_\text{RH}^4}{m_\Phi}\,,
\nonumber
\eea
\noindent
where we used $\rho_\Phi(T_\text{RH})=\rho_R(T_\text{RH})$ and the 
branching ratio is given by
\beq
B_R = \frac{N_\phi\Gamma^{\Phi,d}_{\phi \phi}}{\Gamma^\Phi_{HH}} \simeq
\frac{d^2(25+9\pi^2)}{41472\pi^4}\frac{m_\Phi^8}{\Lambda^8},
\eeq
\noindent
where $N_\phi$ is the number of $\phi$ particles produced per decay, which is $N_\phi=2$ in the present case. The DM relic abundance produced from inflaton decays is thus given by (\ref{Eq:omega})

\beq
\frac{\Omega^{\rm dec}_{\phi} h^2}{0.1} \simeq
6 \times \left( \frac{B_R}{ 10^{-8}} \right)
\left( \frac{T_{\rm RH}}{10^{11}} \right)
\left(\frac{3 \times 10^{13}}{m_\Phi} \right)
\left( \frac{m_\phi}{100}  \right)\,,
\eeq
\noindent
which, for the disformal coupling, can be written

\beq
\frac{\Omega^\text{dec}_{\phi,d} h^2}{0.1}\simeq
 d^2 \left( \frac{T_\text{RH}}{10^{11}}\right)
\left( \frac{m_\Phi}{3 \times 10^{13}} \right)^7
\left(\frac{10^{14}}{\Lambda} \right)^8
\left( \frac{m_\phi}{ 100} \right).
\label{Eq:omegadecayinstant} 
\eeq

\noindent
It is remarkable  that whilst at tree level, one needs to
fine tune tiny dark matter couplings to the inflaton sector
to ensure a branching ratio $B_R \lesssim 10^{-9}$ to avoid
overproduction of dark matter, when one considers radiative production, 
for a BSM scale $\Lambda$ of the order of $10^{14}$ GeV, the disformal
coupling $d$ can easily reach unity without overclosing the Universe.

Moreover, comparing Eqs.~(\ref{Eq:omegatotscat}) 
and (\ref{Eq:omegadecayinstant}),
we see that the production has the same order of suppression in $\Lambda$,
although for a reheating temperature below $\lesssim 10^{12}$ GeV, it is
clear that the radiative decay dominates over the scattering processes. 
To be more precise, we can ask ourselves for which value of $T_\text{RH}$ the scattering rate will begin to produce more dark matter than 
the radiative decay. We obtain
\beq
T_{\rm RH} \gtrsim T_{\rm RH}^\text{eq} = 2 \times 10^{12}\,\text{GeV}.
\label{Eq:trhscat}
\eeq
\noindent
It is remarkable that this temperature does not depend either on 
$m_\phi$ or  $\Lambda$.

\subsection{Conformal production}

It is relevant to compare the disformal production
to the one generated by the conformal coupling of Eq.~(\ref{Eq:conformalcoupling}). It is easy to understand that the part
proportional to $X= \partial^\mu \phi \partial_\mu \phi$ will not be 
very different from the disformal part we just discussed. 
We computed the production 
rate in appendix, Eq.~(\ref{Eq:betacx}), and obtained a value 
of $R(T)=\beta_{c_X} T^{12}/ M_P^8$, with 
$\beta_{c_X} \simeq 10 \, c_X^2$, i.e. with a numerical prefactor of the same order of magnitude as for $\beta_d \simeq 4\, d^2$. Considering the coupling $c_X$ should then give similar
phenomenological results as for the coupling $d$, when $c_X\sim d(M_P/\Lambda)^4$. 
However, the presence of a constant
$c$ term in Eq.~(\ref{Eq:conformalcoupling}) can affect drastically 
the dark matter production. The rate
will then be given by
\beq
R_c(T) \,= \, \beta_c \frac{T^8}{M_P^4}\,,
\eeq
where $\beta_c\simeq 1.1 \times 10^{-2} c^4$. The exact expression is given in Eq.~(\ref{Eq:ratec}). This is computed the in the same manner as $\beta_d$, 
{\it i.e} taking into account {\it all} the Standard Model spectrum
in the initial state. From the production rate $R_c$  we can deduce
the relic abundance after integration on $T$ :
\bea
\frac{\Omega_{\phi,c}^{\rm scat}h^2}{0.1}&&
\simeq 1.6 \times 10^8 \frac{g_0}{g_{RH}^{3/2}}
\sqrt{10}
\frac{\beta_c}{\pi}
\frac{T_{\rm RH}^3}{M_P^3}\,,
\nonumber
\\
&&
\simeq 4.3
\left( \frac{c}{100}\right)^4
\left(\frac{T_{\rm RH}}{10^{11}}\right)^3
\left( \frac{m_\phi}{10^{10}} \right)\,,
\eea
\noindent
for the scattering processes, and

\beq
\frac{\Omega^{\rm dec}_{\phi,c}h^2}{0.1}\simeq
0.7
\left( \frac{c}{100}\right)^4 
\left(\frac{m_\Phi}{3 \times 10^{13}} \right)^3
\left( \frac{m_\phi}{ 10^7}\right)\,,
\eeq
\noindent
for the decaying process, where we used
\beq
\Gamma_{\phi \phi}^{\Phi,c}= \frac{(4 + \pi^2)c^4}{512 \pi^5}
\frac{\mu_\Phi^2 m^3_\Phi}{M_P^4}\,.
\eeq

\noindent
More details regarding the calculations can be found in
the appendix.
We see then that for lower reheating temperature, 
$T_{\rm RH} \lesssim 10^{11}$ GeV, the conformal couplings dominate the dark matter production from scattering 
over the disformal source. That is understandable because the 
dependence on the production rate is lower for conformal coupling 
than disformal coupling. The same can be said concerning the decay 
channel $\Phi \rightarrow \phi \phi$ which dominates for the disformal
coupling. The possibility of having both conformal {\it and} disformal coupling at the same time will be discussed below.

\section{IV. Analysis}

\subsection{Instantaneous reheating case}

\begin{figure}[ht]
\centering
\includegraphics[width=3.in]{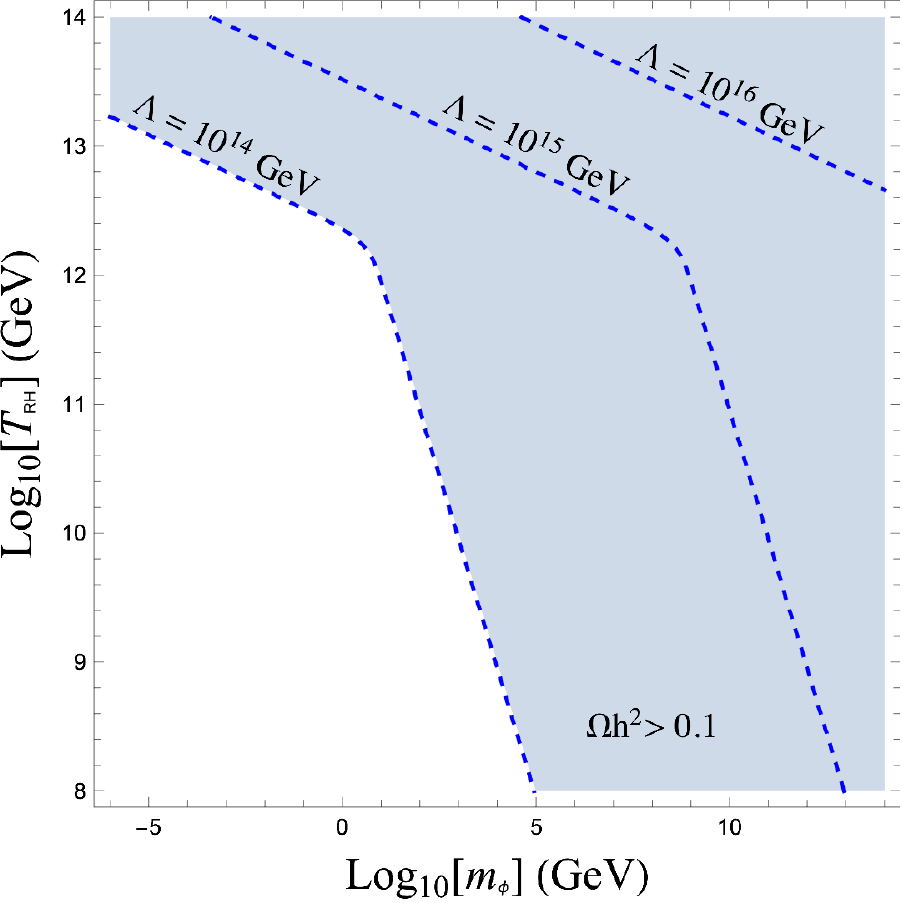}
\caption{\em \small Parameter space allowed in the ($m_\phi$, $T_{\rm RH}$)
plane, for different values of $\Lambda$ for $d=1$ and $c=0$.}
\label{Fig:omega0}
\end{figure}

\noindent
We show in Fig.~\ref{Fig:omega0} the parameter space yielding the correct relic abundance for $d=1$ and $c=0$ in particular the dashed blue curve corresponding to observed dark matter abundance. 
We recognize clearly the two regimes (scattering and decay) from
their different dependence on the reheating temperature, especially the change of regime for $T_\text{RH}=T_\text{RH}^\text{eq} \simeq 2\times 10^{12}$~GeV,
as expected by our approximation (\ref{Eq:trhscat}).
While the scattering process gives a mild dependence
$m_\phi \propto T_{\rm RH}^{1/7}$ 
for $T_{\rm RH} \gtrsim T_{\rm RH}^\text{eq}$, the decay processes implies a harder dependence, $m_\phi \propto T_{\rm RH}^{-1}$.
Notice also that
for BSM scales above GUT scale, $\Lambda \gtrsim 10^{16}$ GeV, it
becomes almost impossible to generate the correct amount of dark matter, 
neither from scattering nor from the inflaton decay, both
processes being too slow to compete with the expansion rate driven
by $H(T)$.

\begin{figure}[ht]
\centering
\includegraphics[width=3.in]{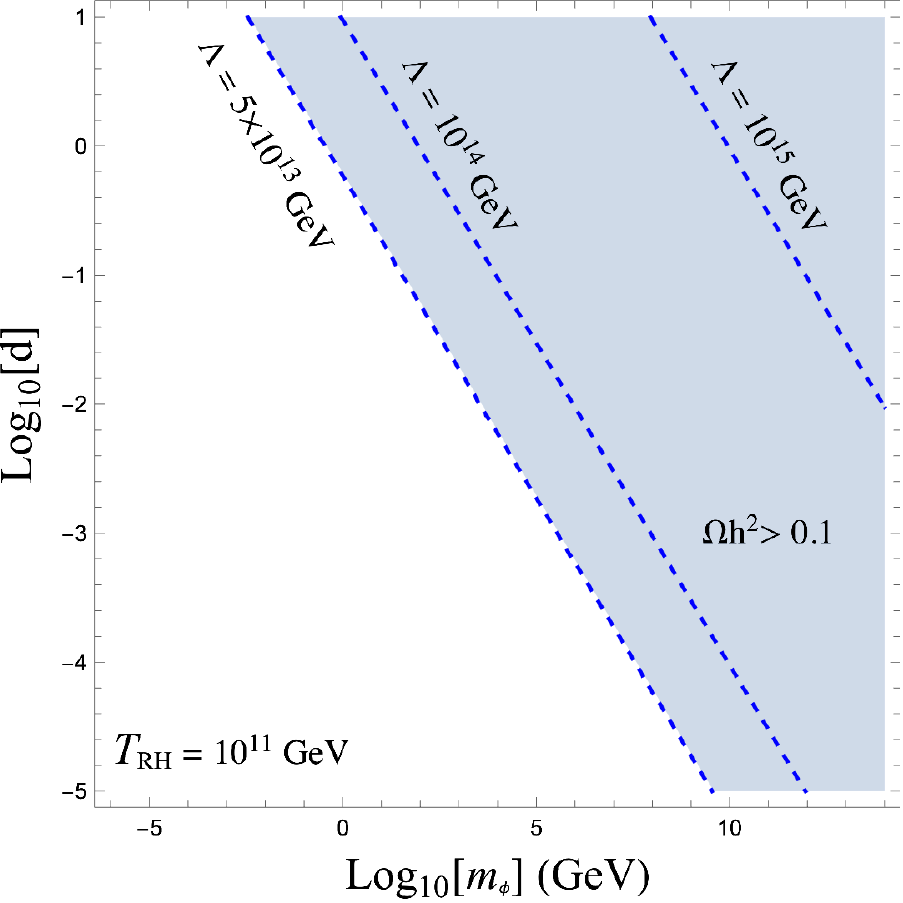}
\caption{\em \small Parameter space allowed in the ($m_\phi$, $d$)
plane, for different values of $\Lambda$ for $T_{\rm RH}=10^{11}$GeV and $c=0$.}
\label{Fig:omegad}
\end{figure}

We also show in Fig.~\ref{Fig:omegad} the allowed region
in the plane ($m_\phi$, $d$)  { assuming disformal couplings only ($c=c_X=0$)} for different values of
$\Lambda$ and $T_{\rm RH} = 10^{11}$ GeV. We observe 
that fairly natural values of $d$, of the order 
of loop factors  $1/(4\pi)^2$, 
make it possible 
to obtain dark matter in sufficient quantity while avoiding overabundance. Still, larger values of $\Lambda$
imposes relatively heavy dark matter, above the TeV-PeV scale to respect the cosmological observations.

\subsection{Non-instantaneous reheating effects}
\label{sec:non-inst}

Until now, we have considered a thermal Universe where reheating took place instantaneously, {\it i.e.} the
energy available in the last oscillations of the inflaton
was instantaneously transferred to the radiative 
bath at $t\simeq (\Gamma^\Phi)^{-1}$,
where $\Gamma^\Phi$ is the decay width of the inflaton. 
However, we  know that non-instantaneous 
perturbative phenonomena can have serious consequences
on the thermal evolution of the Universe~\cite{Garcia:2017tuj}, and even 
more on the production of dark matter in its first
instants if the modes of production show a large
dependence on the energy of the processes involves~\cite{Garcia:2020eof},  is the case for the disformal
(conformal) scenario with a rate 
$R(T)\propto T^{12}$ ($T^8$) respectively.

More generally, 
the dark matter production during the reheating may not be negligible, especially when $R(T)\propto T^{n+6}$ with $n\geq6$, due to the effects of non-instantaneous reheating~\cite{Garcia:2017tuj}, non-instantaneous thermalization in~\cite{Reheating}, and non-quadratic inflaton potentials during the reheating stage~\cite{Garcia:2020eof}.
In our case the contributions from the conformal and disformal coupling terms proportional to $c_X$ and $d$, respectively, correspond to $n=6$, whereas the reaction rate of the conformal coupling term proportional to $c$ corresponds to $n=2$. The interference term corresponds
to $n=4$.
In this section we discuss the non-instantaneous reheating effect, while assuming instantaneous thermalization and no preheating contributions.

As a specific example, we consider the Starobinsky model for inflation~\cite{Starobinsky:1980te} where the inflaton oscillation is described by $V(\Phi)=\frac{1}{2} m_\Phi^2\Phi^2$ after the end of inflation.
Then, we may use the result for the enhancement of the DM production discussed in Ref.~\cite{Anastasopoulos:2020gbu}. Solving the 
complete set of combined equations for the inflaton density 
$\rho_\phi$, the radiation density $\rho_R$ and the dark matter
has been carried out and analyzed in~\cite{Garcia:2017tuj} for 
any kind of dark matter production cross section and more recently in 
\cite{Garcia:2020eof} for any type of inflationary potential.
To summarize these works, we just need to understand that the reheating 
process being non-instantaneous, the temperature of the primordial plasma evolves from a null temperature to a maximum value $T_{\rm max}$
before decreasing until the radiation density $\rho_R$ catches
the inflaton density $\rho_\phi$, defining the thermal era, happening
at the reheating temperature $T_{\rm RH}$. 
The evolution between $T_{\rm max}$
and $T_{\rm RH}$ is rather complex, but the main point
is that the production of dark matter for cross-sections with a large
temperature dependence, of the order $T^{n+6}$ with $n \geq 6$, 
is largely affected by the maximal
temperature as most of the dark matter
is produced
at this instant. In comparison with an instantaneous treatment, 
there is a boost factor which is a function of $T_{\rm max}/T_{\rm RH}$. We summarize the results in the following paragraph.

The maximal temperature $T_{\rm max}$ and $T_\text{RH}$ are obtained as
\bea
T_{\rm max} &=& \left(\frac{45}{32}\frac{3^{1/10}}{2^{4/5}}\frac{y^2 m_\Phi M_P \rho_{\rm end}^{1/2}}{g_*(T_{\rm max}) \pi^3}\right)^{1/4}\\
&\simeq& 1.6\times10^{13}~{\rm GeV}\times \left(\frac{106.75}{g_*(T_{\rm max})}\right)^{1/4}\nonumber\\
&&\times \left(\frac{\mu_\Phi}{10^{10}~{\rm GeV}}\right)^{1/2}\left(\frac{\rho_{\rm end}}{0.175m_\Phi^2M_P^2}\right)^{1/8},\nonumber\\
T_\text{RH} &=& \left(\frac{9}{40}\frac{y^4 m_\Phi^2 M_P^2}{g_\text{RH}\pi^4}\right)^{1/4}
\label{Eq:tRH}
\\
&\simeq&1.9\times10^{11}~{\rm GeV}\times  \left(\frac{106.75}{g_\text{RH}}\right)^{1/4}\nonumber\\
&&\times\left(\frac{\mu_\Phi}{10^{10}~{\rm GeV}}\right)\left(\frac{3\times10^{13}~{\rm GeV}}{m_\Phi}\right)^{1/2}\nonumber
\eea

\noindent
where again $T_\text{RH}$ is defined by $\rho_\Phi(T_\text{RH})=\rho_R(T_\text{RH})$, and we assume $g_*(T_{\rm max})=g_*(T_\text{RH})=g_\text{RH}$ in the following analysis.
We have used the inflaton decay width $\Gamma^\Phi_{HH}\equiv y^2m_\Phi/8\pi$ with $y\equiv \mu_\Phi/m_\Phi$ from Eq.~(\ref{Eq:inflaton-higgs}), where $m_\Phi^2\simeq 24\pi^2A_{S^*}M_P^2/N_*^2$ with $\ln(10^{10}A_{S^*})=3.044$~\cite{planck,Akrami:2018odb} and $N_*\simeq 55+0.33\ln y$~\cite{Anastasopoulos:2020gbu}.
Then, for $n=6$, we obtain the boost factor $B^{\rm scatt}\equiv n_\phi^{\rm non-inst}(T_\text{RH})/n_\phi(T_\text{RH})$ given by
 \bea
B^{\rm scat}=f\frac{56}{3}\log\frac{T_{\rm max}}{T_\text{RH}},
\label{Eq:boostfactor}
\eea

\noindent
where $f\simeq 1.2$ to match the numerical results.
Notice that for $n=2$, which is the case of $d=c_X=0$ with $c\neq0$, we do not have such an enhancement, since the DM production is dominated at $T_{\rm RH}$.

\begin{figure}[ht]
\centering
\includegraphics[width=3.in]{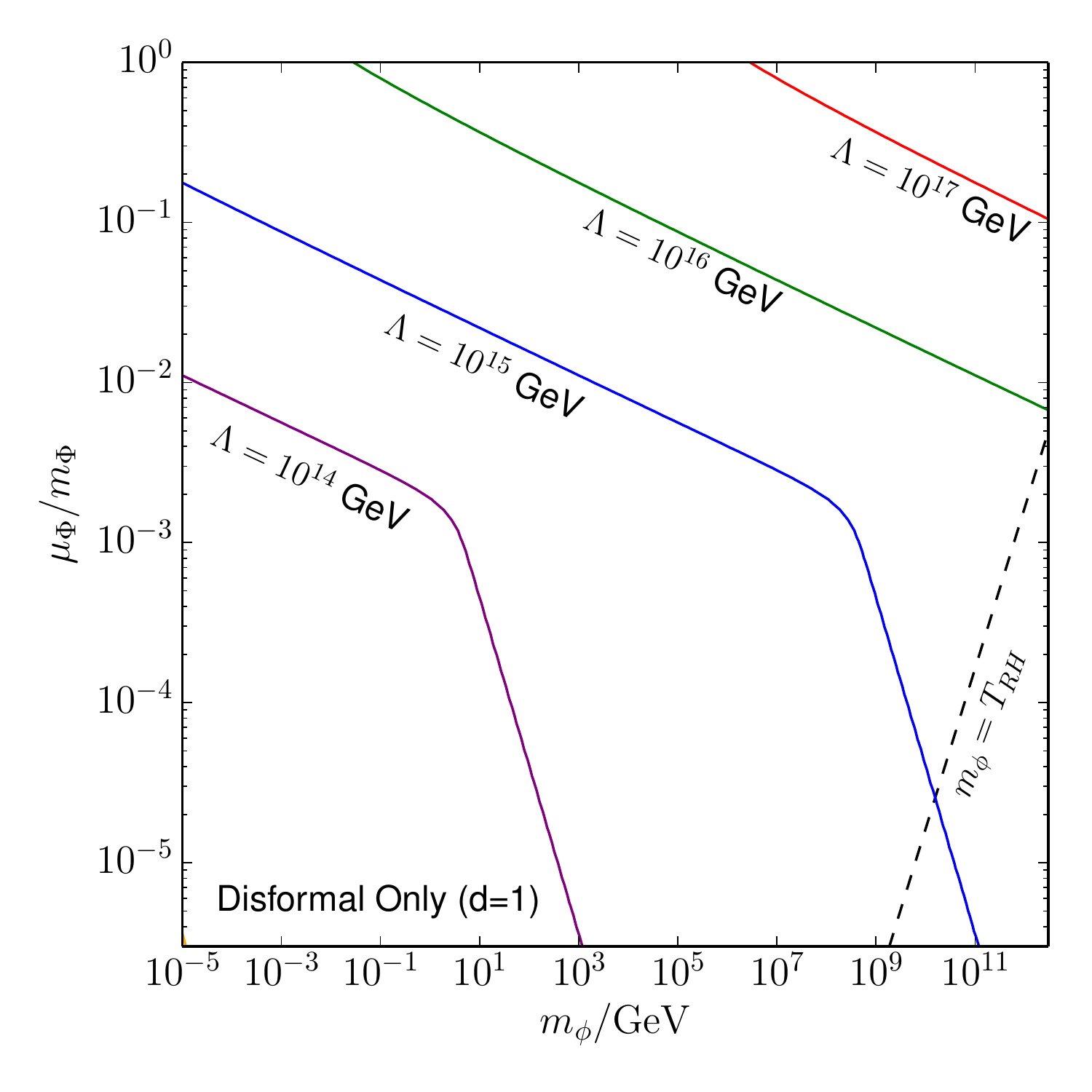}
\caption{\em \small Parameter space allowed in the ($m_\phi$, $\mu_\Phi/m_\Phi$)
plane, for different values of $\Lambda$ for $d=1$ and $c=c_X=0$.}
\label{Fig:omegad1}
\end{figure}

Figure~\ref{Fig:omegad1} shows the contours of $\Omega^{\rm scat}_\phi h^2+\Omega^{\rm dec}_\phi h^2=0.1$ where only the disformal coupling contributes, namely $c=c_X=0$ and $d\neq0$, and we take $d=1$, taking
into account the effect of non-instantaneous reheating just discussed above.
Notice that in the bottom-right corner of the figure, the dark matter mass is in excess of $T_{\rm RH}$, and thus the scattering contributions get further suppressed, which is however irrelevant for smaller $\Lambda$ ($\lesssim 10^{15}$ GeV), since the decay contribution dominates in that parameter space,  the domination
occuring from Eqs.(\ref{Eq:trhscat}) and (\ref{Eq:tRH}) for $\mu_\Phi \simeq 3\times 10^{-3}~m_\Phi$. It would be interesting, in this framework, to compare, the disformal to the conformal production of dark matter.

\subsection{Combining conformal and disformal coupling}

The conformal and disformal couplings may coexist.
For instance, we can take both $d$ and $c$ being nonzero, while $c_X=0$.
\noindent
Incorporating nonzero contribution from the $c$ coupling, we obtain the scattering reaction rate
\bea
R(T) = \beta_c \frac{T^8}{M_P^4} + \beta_{cd}\frac{T^{10}}{\Lambda^4 M_P^2} + \beta_d \frac{T^{12}}{\Lambda^8}
\label{Eq:totalratecd}
\eea
where $\beta_c \simeq 1.1 \times 10^{-2}\, c^4$ is given in Eq.~(\ref{Eq:ratec}) and $\beta_d \simeq  4 \,d^2$ is given in Eq.~(\ref{eq:betad}). The quantity $\beta_{cd}$, arising from interferences between conformal and disformal, couplings is given by 
\bea
\beta_{cd} &=& -c^2\,d\,\frac{24\zeta(5)^2}{\pi^5}\simeq-8.4 \times10^{-2}\,c^2\,d \,.
\eea

\begin{figure}[ht]
\centering
\includegraphics[width=3.in]{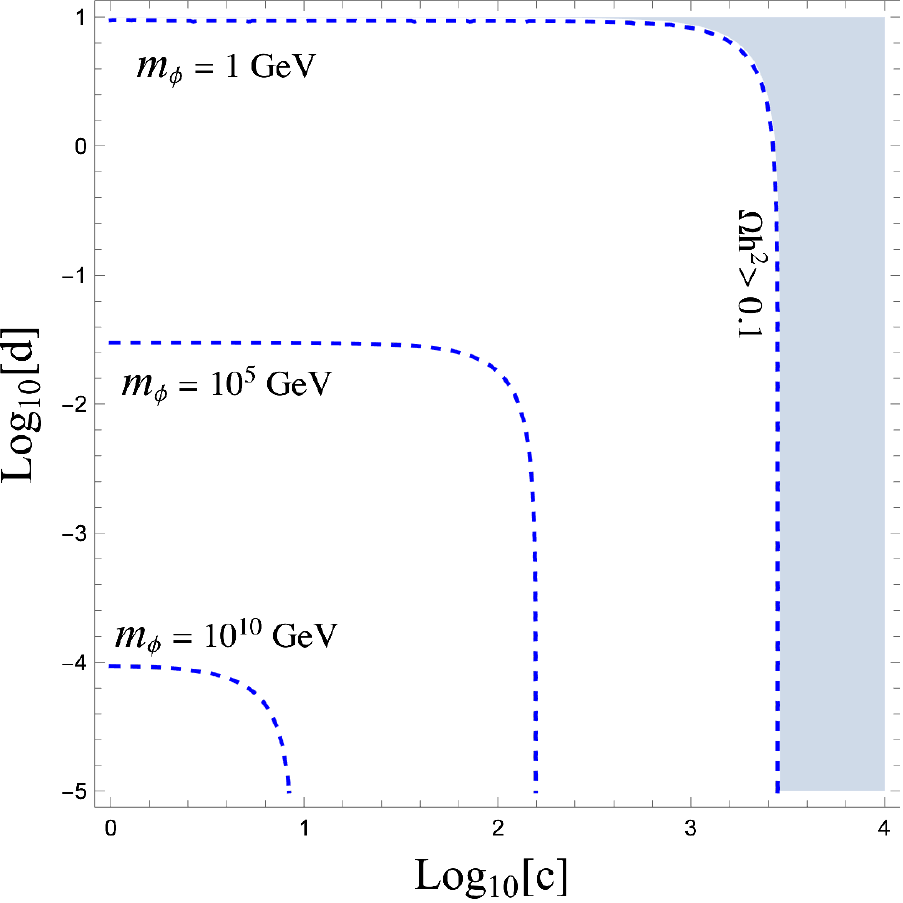}
\caption{\em \small Parameter space allowed by cosmological constraint
in the plane ($c$, $d$) for different dark matter masses $m_\phi$ and $T_{\rm RH}=10^{11}$ GeV and $\Lambda=10^{14}$ GeV.}
\label{Fig:omegamaster}
\end{figure}

The radiative inflaton decay is also affected by the conformal coupling contributions, and thus we obtain
\begin{multline}
\Gamma^\Phi_{\phi\phi}=
\frac{d^2\mu_\Phi^2 m_\Phi^7}{512\pi^5\Lambda^8}
\left[\left(\frac{5}{36}+\frac{2c^2\Lambda^4}{dm_\Phi^2M_P^2}\right)^2 \right. \\
\left. +\pi^2\left(\frac{1}{12}+\frac{c^2\Lambda^4}{dm_\Phi^2M_P^2}\right)^2
\right]\,,
\label{Eq:totaldecay}
\end{multline}
giving a branching ratio to a DM pair of
\begin{multline}
    B_R= 
\frac{d^2 m_\Phi^8}{32\pi^4\Lambda^8}
\left[\left(\frac{5}{36}+\frac{2c^2\Lambda^4}{dm_\Phi^2M_P^2}\right)^2 \right. \\
\left. +\pi^2\left(\frac{1}{12}+\frac{c^2\Lambda^4}{dm_\Phi^2M_P^2}\right)^2
\right] \,.
\label{Eq:totalBR}
\end{multline}
\noindent
Combining the relic abundance produced by scattering integrating Eq.~(\ref{Eq:boltzmann0}) \footnote{ By taking the non-instantaneous reheating into account, one obtains
\bea
\frac{\diff}{\diff T}(n_\phi T^{-8})=-\frac{8}{3}\frac{R(T)}{HT^9}
\eea
with $H(T)=(2/5)\Gamma^\Phi_{HH}(T/T_{\rm RH})^4$, instead of~Eq.~(\ref{Eq:boltzmann}).
} with the 
rate~(\ref{Eq:totalratecd}),
combined with the boost factor due to non-instantaneous
thermalization~(\ref{Eq:boostfactor}) and adding
the decay process~(\ref{Eq:totaldecay}) we obtain
\begin{widetext}
\bea
&&
\frac{\Omega^{\rm tot}_\phi h^2}{0.1}=
\nonumber
\\
&&
9.1\times10^{-10}d^2\left(\frac{T_{\rm RH}}{10^{11}}\right)^3\left[
5.9\ln\left(\frac{T_{\rm max}}{T_{\rm RH}} \right)
\left(\frac{T_{\rm RH}}{10^{11}} \right)^4
\left(\frac{10^{14}}{\Lambda} \right)^8
+\left( \frac{c}{100\sqrt{d}} \right)^4
-1.2\times10^{-4} \frac{c^2}{d}
\left(\frac{T_{\rm RH}}{10^{11}} \right)^2
\left(\frac{10^{14}}{\Lambda} \right)^4
\right]\left(\frac{m_\phi}{\rm GeV}\right)
\nonumber
\\
&&
+4.6\times10^{-3} d^2 \left( \frac{T_{\rm RH}}{10^{11}}\right)
\left(\frac{10^{14}}{\Lambda} \right)^8
\left(\frac{m_\Phi}{3 \times 10^{13}} \right)^7
\left[ 
\left(1+\frac{72}{5}\frac{c^2\Lambda^4}{d m_\Phi^2 M_P^2} \right)^2
+\pi^2\left(\frac{3}{5} + \frac{36}{5} \frac{c^2\Lambda^4}{d m_\Phi^2 M_P^2}\right)^2
\right]\left(\frac{m_\phi}{\rm GeV} \right),
 \label{Eq:masterequation}
\eea
where we used the following results (see, for instance, Ref.~\cite{Garcia:2020eof})
\begin{equation}
\frac{n^{\rm scat}_\phi(T_\text{RH})}{T_{\rm RH}^3}\simeq \sqrt{\frac{90}{g_{\rm RH}\pi^2}}\left[\frac{2}{3}\beta_c\frac{T_{\rm RH}^3}{M_P^3}+\frac{4}{3}\beta_{cd}\frac{T_{\rm RH}^5}{M_P\Lambda^4}+\frac{1}{7}B^{\rm scat}\beta_d\frac{M_PT_{\rm RH}^7}{\Lambda^8}\right]\,, \quad \text{and} \quad ~
\frac{n^{\rm dec}_\phi(T_\text{RH})}{T_{\rm RH}^3}\simeq \frac{g_{\rm RH}\pi^2}{18}B_R\frac{T_{\rm RH}}{m_\Phi}\,,
\end{equation}
\end{widetext}
\noindent
which is the main result of our work. Eq.~(\ref{Eq:masterequation})
gives the total amount of dark matter produced in a model with a
combination of disformal ($d$) and conformal ($c$) couplings, taking into account
production through scattering from the thermal bath {\it and}
radiative decay of the inflaton, together with instantaneous effects due its the finite width. We illustrate our results in Fig.~\ref{Fig:omegamaster} where we plot the region of the parameter
space allowed in the plane ($c$, $d$) for different dark matter masses
$m_\phi$, fixing $T_{\rm RH}=10^{11}$ GeV and $\Lambda=10^{14}$ GeV.
We clearly distinguish the two regimes, and for which values of $c$
the conformal couplings begin to dominate over the disformal one.
For our choice of parameters, the decay rate dominates
the production of $\phi$ in Eq.~(\ref{Eq:masterequation}). It is interesting
to notice that for any dark matter mass, there exists a point in the parameter space, with reasonable value of $c$ and $d$, respecting the
cosmological constraint despite the large suppression due to high
BSM physics scales.

\begin{figure}[ht]
\centering
\includegraphics[width=3.in]{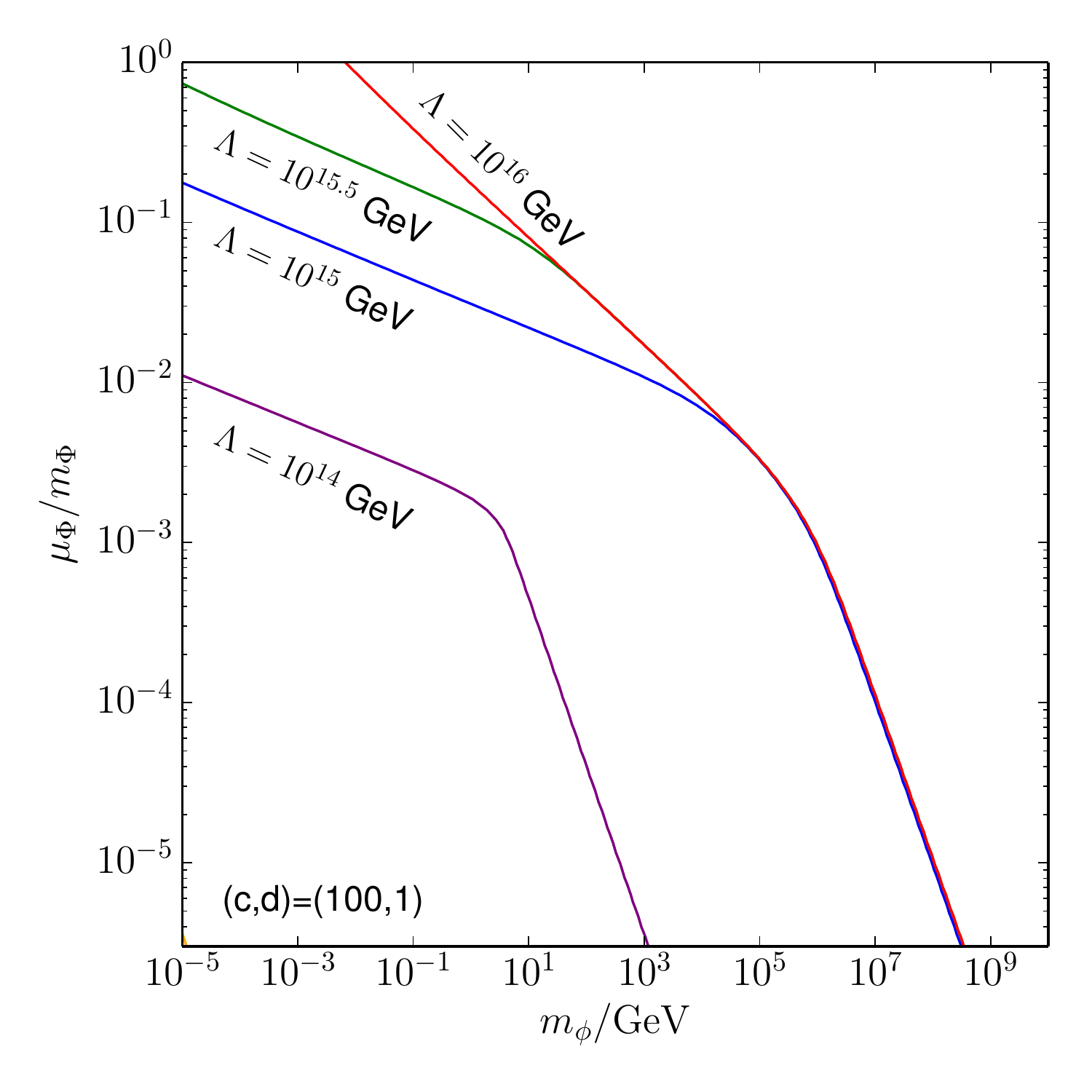}
\caption{\em \small Parameter space allowed in the ($m_\phi$, $\mu_\Phi/m_\Phi$)
plane, for different values of $\Lambda$ for $c=100$ and $d=1$.}
\label{Fig:omegad1c1e2}
\end{figure}

If one looks into more details at the zone of influence of the disformal 
coupling versus the conformal one, we find that for
\begin{equation}
\Lambda\simeq 10^{15}~{\rm GeV}\left(\frac{10^4d}{c^2}\right)^{1/4}\left(\frac{T_{\rm RH}}{4\times10^{12}}\right)^{1/2}\left(\frac{B^{\rm scatt}}{74}\right)^{1/8}\,,
\end{equation}
\noindent
both processes gives a similar contribution to the relic abundance, smaller
values of $\Lambda$ favouring of course the disformal production.
We illustrate this situation in Fig. \ref{Fig:omegad1c1e2} where
we take $c=100$ and $d=1$. For $\Lambda$ below $\sim 10^{15}$ GeV we recognize the characteristic of disformal production observed in Fig.(\ref{Fig:omegad1}) whereas
for $\Lambda \gtrsim 10^{15}$ GeV, the production begins to be independent of $\Lambda$,
which is a clear signature of a conformal production of dark matter.

\section{V. Conclusions}

We have shown that in models where the {\it geometrical} metric governing gravitational physics is different
from the {\it dynamical} metric felt by the standard model particles via a scalar field $\phi$, this scalar $\phi$ can play the role of  dark matter. In this scenario,  dark matter is produced via freeze-in and it is possible to respect cosmological constraints on the relic abundance of dark matter. Moreover, this can happen even though the suppression scale of the coupling between the scalar field and matter particles could be almost as large as the Planck scale. The number density of scalar dark matter particles would then be produced in the early stages of the Universe, by a freeze-in mechanism due to its very feeble coupling to the Standard Model sector, {i.e. due to the large suppression scale. 
In such a disformal dark matter scenario where the dark matter field would be disformally coupled to the standard model fields,
the low production rate of $\phi$ would  be counterbalanced by a large mass $m_\phi$, making disformally coupled scalars
 perfect candidates for experiments looking for superheavy dark components like
ANITA or IceCube~\cite{Heurtier:2019git}.

\vskip.1in
{\bf Acknowledgments:}
\noindent 
The authors want to thank especially E. Babichev, C. Charmousis, 
E. Dudas and
Marcos A. G. Garcia for very  insightful
discussions. This work was supported in part by the France-US PICS MicroDark. 
The work of MP was supported by the Spanish Agencia Estatal de Investigaci\'{o}n through the grants FPA2015-65929-P (MINECO/FEDER, UE),  PGC2018-095161-B-I00, IFT Centro de Excelencia Severo Ochoa SEV-2016-0597, and Red Consolider MultiDark FPA2017-90566-REDC. This project has received funding/support from the European Unions Horizon 2020 research and
innovation programme under the Marie Skodowska-Curie grant agreements Elusives ITN No. 674896
and InvisiblesPlus RISE No. 690575. 
The work of KK was supported by a KIAS Individual Grant (Grant No. PG080301) at 
Korea Institute for Advanced Study.

\section*{Appendix}

\section{A. Production rate:  definitions}

Assuming that DM is predominantly produced by $2\rightarrow 2$ annihilations of SM particles, the Boltzmann equation for the DM number density can be written as
\begin{align}
    \dfrac{\diff n_\text{DM}}{\diff t}+3Hn_\text{DM}\,= \,R(T)\,,
\end{align}
where the quantity on the right-hand-side $R(T)$ represents the temperature-dependent DM production rate per unit of volume and time. The rate can be expressed as a sum of the contribution of SM species of spin $i$ to  
\begin{equation}
    R(T)\,=\,\sum_{i=0,1/2,1} N_i R_i\,= 4 R_0 + 45 R_{1/2}+12 R_1\,,
\end{equation}
where $N_i$ is the number of the SM species of spin $i$. The partial rate $R_i$ can be expressed as
\begin{widetext}
\begin{equation}
   R_{i}(T)\,=\, \frac{1}{1024 \pi^6}\int f_i(E_1) f_i(E_2) E_1 \diff E_1 E_2 \diff E_2 \diff \cos \theta_{12}\int |{\cal M}_i|^2 \diff \Omega_{13}\,,
   \label{eq:def_rate_integral}
\end{equation}
\end{widetext}
with $p_j(E_j)$ is the 4-momentum (energy) of particles $j=1,2,3,4$ for processes $1+2\rightarrow 3+4$ with 1, 2 being particles of the SM and 3, 4 dark matter states. $f_i$ represent the Bose-Einstein ($i=0,1$) and Fermi-Dirac ($i=1/2$) statistics distribution functions. $\theta_{13}$ and $\theta_{12}$ are the angle formed by momenta of 1,3 and 1,2 respectively. The differential solid angle can be expressed as $\diff \Omega_{13}=2 \pi \diff \cos \theta_{13}$. These kinematics quantities are related to the Mandelstam variables in the ultra-relativistic limit $t= (s/2)( \cos \theta_{13}-1)$ and $  s = 2E_1E_2(1-\cos \theta_{12})$. More details can be found in the Appendices of Ref.~\cite{Anastasopoulos:2020gbu}.

\section{B. Production rate:  scattering}
\label{sec:productionrate}

\subsection{Rate for a generic amplitude}
Assuming an amplitude squared for the process $i+i\rightarrow \text{DM+DM}$, where $i$ denotes one SM particle of spin $i$, of the form\footnote{As $s+t+u=0$ in the ultrarelativistic limit, our expression 
contains all the possible processes.}.
\begin{equation}
 |\mathcal{M}_{i}|^2=\sum_{n,k=0} c_{nk}^i
 \frac{s^{n}t^k}{\Lambda^{2(n+k)}}\,,
 \label{eq:ampgeneric_spini}
\end{equation}
the integrated amplitude squared reads
\begin{equation}
    \int \diff \Omega_{13} |\mathcal{M}_{i}|^2= \sum_{n,k=0} c_{nk}^i (-1)^k \frac{4 \pi  }{k+1} \dfrac{s^{n+k}}{\Lambda ^{2(n+k)}}\,.
\end{equation}
Taking the integral expression of Eq.~(\ref{eq:def_rate_integral}), the contribution of a particle of spin $i$ to the rate is
\begin{widetext}
\begin{equation}
    R_{i}(T)\,=\,\sum_{n,k=0} c_{nk}^i  (-1)^k\frac{2^{2(n+k)}  \Gamma^2(n+k+2) \zeta^2(n+k+2) T^{2(n+k+2)}}{128 \pi ^5 (k+1)(n+k+1) \Lambda ^{2(n+k)}} \times  \begin{cases}
1\,,  & \,(i=0,1)\,,\\
\left(1-2^{-(1+n+k)}\right)^2\,, &\,(i=1/2)\,,
\end{cases}
\end{equation}
and the corresponding contribution to the relic density is given by
\begin{equation}
   \Omega_\text{DM}^i h^2 \simeq \sum_{n,k} c_{nk}^i (-1)^k 
   \frac{135 \sqrt{10} M_P m_{\text{DM}}}{  256 \pi ^8 g_*^{3/2}}
   \frac{2^{2(n+k)}    \Gamma^2(n+k+2) \zeta^2(n+k+2) T_\text{RH}^{2(n+k)-1}}{ (k+1)(n+k+1) (2(n+k)-1) \Lambda^{2(n+k)}} \dfrac{s_0 h^2}{\rho^0_c}\times   \begin{cases}
1\,,  \\
\left(1-2^{-(1+n+k)}\right)^2\,,
\end{cases}
\end{equation}
\end{widetext}
with $m_{\rm DM}$ being the dark matter mass, where the first and second cases correspond respectively to the  Bose-Einstein ($i=0,1$) and the Fermi-Dirac ($i=1/2$) statistics for the initial state particles. We used the expression of the Hubble rate in terms of the SM temperature in the radiation domination era $H(T)=(g_* \pi^2/90)^{1/2}T^2/M_P$ and considered constant relativistic degrees of freedom for simplicity $g_*=g_{\text{RH}}$. $s_0$ and $\rho_c^0$ are the entropy density and critical density of the present time. Assuming that for each SM particle of spin $i$, the DM production amplitude squared is given by Eq.~(\ref{eq:ampgeneric_spini}), the total relic density can be expressed as
\begin{equation}
   \Omega_\text{DM} h^2\, = \, 4\, \Omega_\text{DM}^0 h^2+45\, \Omega_\text{DM}^{1/2} h^2+12\,\Omega_\text{DM}^1 h^2\,.
\end{equation}
\subsection{Rate for disformal couplings}
The amplitudes $ |\mathcal{M}_i|^2$ for the processes  $i+i\rightarrow \text{DM+DM}$, where $i$ denotes one SM particle of spin $i$, are given by 
\begin{equation}
 |\mathcal{M}_0|^2\,=\,d^2\dfrac{t^2(s+t)^2}
 {8 \Lambda^8}\,,
\end{equation}

\begin{equation}
 |\mathcal{M}_{1/2}|^2\,=\,-d^2\dfrac{t(s+t)(s+2t)^2}
 {16 \Lambda^8}\,,
\end{equation}

\begin{equation}
 |\mathcal{M}_{1}|^2\,=\,d^2\dfrac{t^2(s+t)^2}
 {4 \Lambda^8} \,.
\end{equation}
The total rate is given by
\begin{equation}
    R_d(T)= d^2 \dfrac{100589 \pi^7}{76204800} \dfrac{T^{12}}{\Lambda^8} \equiv  \beta_d \dfrac{T^{12}}{\Lambda^8}\,,
    \label{eq:betad}
\end{equation}
with $\beta_d \simeq 4 d^2$.
\subsection{Rate for conformal couplings}

As previously mentioned, in this case only the scalar particles contribute to the rate. The amplitude is given by

\begin{equation}
 |\mathcal{M}_0|^2=
 \frac{s^2}{8 M_P^4} \left(c_X\frac{ s}{M_P^2}-2 c^2\right)^2\,,
\end{equation}

\noindent
For the case $c=0$, the total rate is given by
\begin{equation}
    R_{c_X}(T)= c_X^2 \dfrac{64 \pi^7}{19845} \dfrac{T^{12}}{M_P^8} \equiv  \beta_{c_X} \dfrac{T^{12}}{M_P^8} \,,
    \label{Eq:betacx}
\end{equation}

\noindent
with $\beta_{c_X} \simeq 9.74 c_X^2$. For the case $c_X=0$, the total rate is given by

\begin{equation}
    R_c(T)= c^4 \dfrac{ \pi^3}{2700} \dfrac{T^{8}}{M_P^4} \equiv  \beta_c \dfrac{T^{8}}{M_P^4} \,,
    \label{Eq:ratec}
\end{equation}

\noindent
with  $\beta_c \simeq  1.1 \times 10^{-2} c^4 $.

\vspace{-.5cm}
\bibliographystyle{apsrev4-1}

\end{document}